\documentstyle[sprocl]{article}

\def\nostrocostrutto#1\over#2{\mathrel{\mathop{\kern 0pt \rlap 
  {\raise.2ex\hbox{$#1$}}}
  \lower.9ex\hbox{\kern-.190em $#2$}}}
\newcommand{\be}{\begin{equation}}
\newcommand{\ee}{\end{equation}}
\newcommand{\ba}{\begin{eqnarray}}
\newcommand{\ea}{\end{eqnarray}}

\newcommand{\eref}[1]{(\ref{#1})}

\begin{document}
\pagestyle{empty}

\title{CLAN CONCEPT IN  MULTIPARTICLE DYNAMICS  AND THE NB "ENIGMA"} 

\author{ALBERTO GIOVANNINI} 

\address{Dipartimento di 
Fisica Teorica -- Universit\`a di Torino and \\ 
INFN -- Sez. di Torino, Via P. Giuria 1,  10125 Torino -- Italy} 

\maketitle\abstracts{              
A summary of main results on NB regularity  and related
clan concept since their first appearance
in multiparticle dynamics is presented.} 

\section{Introduction } 

The main scope of this short contribution is to call  the
attention  on the impact of Negative 
Binomial (NB) regularity in our field as documented by
the  collection of Proceedings of  
International Symposia on Multiparticle Dynamics since 1972.  
It is  also quite instructive to follow the evolution 
of the interpretation  of NB regularity    
from its first appearance in cosmic
ray physics in the sixties and in the accelerators region in the
seventies, up to its discovery in the eighties in limited regions 
of phase space
in the CERN $p\bar p$ energy domain, and then in the nineties 
 in $e^+e^-$ annihilation at LEP in
samples of events of different topologies up to the very recent 
suggestion presented at this Symposium to associate 
NB Multiplicity Distribution (MD) to quark jets of different flavour. 
This perspective is motivated by two facts.
First, NB regularity in $e^+e^-$ annihilation seems to be better
satisfied the more fundamental and elementary is the chosen level 
of  investigation of multiparticle final states: NB regularity is 
less approximate  for the MD of
the most primitive object occurring  in Multiparticle Dynamics, 
i.e., the single jet of fixed flavour, and becomes
  more and more approximate in going from single jet to sample of two jet 
 events and finally   to the  full sample of events.
Second, it should be noticed that observed approximations of the 
regularity at a certain level of investigation 
can be in part  removed  by  properly weighting  NB MD's 
occurring with an  higher degree of 
accuracy  in   the elementary  substructures  present at a more
primitive level.
It is clear that if these facts would be confirmed also in other classes
of collisions it would be quite natural to consider the single jet 
structure as an effective building block in multiparticle production 
as it is now in $e^+e^-$ annihilation, and 
the study of more
complex final states structures  as the result of  
compositions
of single jets whose evolution  is  controlled by NB MD's
and related clan structure properties.
These remarks give a deep and striking physical meaning to the  1979 
discovery  that
 QCD jets  can be described in LLA in terms of Markov branching processes 
of NB type  dominated by gluon
self interaction, the non-linear QCD vertex,  and to the
subsequent observation that clans at parton level are bremsstrahlung
gluon jets. Accordingly, 
 the statement that QCD parton shower formation 
is the main dynamical mechanism responsible for multiparticle production
turns out to be  a consequence of the intimate fractal structure of
jets proposed in our field - once again - since 1979. 
Along this line of thought an integrated description of multiparticle 
correlations and MD's is possible. In addition generalized hadron 
parton duality as formulated in the NB framework
 could be indeed   a highly simplified 
approximated     hadronization prescription and opens a new
interesting research line on hadronization models.

\section{Facts}

\subsection{Prehistory (1966)}
NB regularity is discovered in cosmic ray physics\cite{wolf}.  
NB MD  is called Polya-Eggenberger distribution by the name of 
the scientists who applied it to the diffusion of contagious
diseases in 1926.
Fluctuations of the pionization component in hadron showers originated by
a primary  hadron at different primary energies $E_0$ are well described
by NB MD's.

Violations of  multiperipheral model predictions are clearly
visible starting from $E_0 \simeq$  30 GeV. 
Interestingly the Authors suggest
the analytical structure of standard NB parameters  in terms of the 
primary energy $E_0$,  i.e., the average charged particle multiplicity 
$\bar n$ and the parameter $k$, which is linked to the dispersion $D$ by  
$D^2 = \bar n + \bar n^2/k$,  turn out to be respectively
\be
\bar n(E_0) \sim 1.8 (E_0)^{1/4}
\ee
and
\be
k(E_0) \sim 0.4 (1  -  e^{-1.8~10^{-4} E_0})
\ee
It follows that  at $E_0$ = 30 GeV, $\bar  n \sim$ 4 and $k \sim$ 465, 
whereas at $E_0 = 10^4$ GeV $\bar n \sim$ 18 and $k \sim$ 3,  i.e.,  
$k$ is a decreasing
function of the primary energy  as $\bar n$  increases.
A large part of the story of NB regularity in {\it full phase space} is 
in a certain sense  already contained in the just mentioned results. 
 Very large  $k$ values suggest that the MD can be
safely approximated by  multiperipheral model
predictions  (a Poisson MD)  and the onset of correlations 
among produced particles
is controlled by the increasing of $1/k$ at very large primary
energies. See also the proceedings of the Wuhan 1991 symposium 
for an extensive discussion on the subject.

\subsection{NB regularity in the accelerator region  (Zakopane 1972, 
Pavia 1973) } 
NB (Polya-Eggenberger) MD is  suggested  for describing n-particle MD
in full phase space in a simple  model of multiparticle
production which was
motivated by the attempt to describe observed deviations  from
multiperipheral model predictions in the accelerator region\cite{agcim}. 
NB MD is discussed at the Zakopane Symposium in June 1972.
Among other interesting features it is pointed out that the new
distribution obeys in a certain limit KNO scaling and that
related n-particle correlation functions depend on 
two-particle correlation functions only. The explicit
formula and its derivation are also given.
Subsequent systematic analysis of final charged MD's in all
available experiments in the accelerator region confirmed that NB MD 
is a very good candidate to describe MD in full phase space\cite{aglett}.
The main results of this accurate analysis have been presented
at the Pavia Symposium in 1973. 
The analysis has been completed
in 1977\cite{77paper}. 53 experiments were examined
in total.
The general trend of NB parameters is the same observed in
cosmic ray physics, i.e., $\bar n$ and $1/k$ are growing as $p_{lab}$ 
increases in all reactions involving protons on hadronic
targets. In addition, at lower $p_{lab}$ $1/k$ becomes negative
after crossing the  zero at approximately 30 GeV/c.

\subsection{NB regularity  in symmetric (pseudo)-rapidity intervals 
in $p\bar p$ energy domain and  
then in all classes of reactions (Kyriat 1985, La Thuile 1989)} 

NB regularity is rediscovered by UA5 collaboration  and
extended  to symmetric pseudorapidity intervals in $p\bar p$ 
collisions at c.m. energy $\sqrt{s}$ = 200 GeV, 546 GeV
and 900 GeV.
First results are presented by G. Ekspong at the Kyriat
Symposium in 1985.
After the impressive work by NA22, EMC, HRS and NA5 collaborations -
so well summarized by N. Schmitz in his talk at the La Thuile
Symposium in 1989 - the new statement is: 
MD's in all classes of collisions in full phase space and in
symmetric rapidity intervals are quite well  fitted  by NB MD.
One problem  shows up in $p\bar p$ at 900 GeV. NB  does not
fit data so well in the largest rapidity intervals.
A  shoulder is visible in the experimental distribution.

\subsection{Clan concept (Seewinkel  1987)} 

Clan concept has been introduced  in order to  interpret
the wide occurrence of approximate NB regularity in MP production.
A clan is a group of  particles of common ancestor (Sippe in 
German language).
Each clan contains at least one particle (by definition).
Clans are independently produced (by assumption).
This new concept is discussed at the 1986 Seewinkel Symposium\cite{see}
 and extensively
applied to experimental data in a series of papers by the same
authors\cite{app}.  
The  distribution of particles in an average clan is logarithmic  (or
a weighted geometric)  and its composition with independently emitted 
(i.e.  Poissonian) clans leads to the observed NB MD.
\be
G_{NB}(\bar n,k;z)= \left[ 1- \frac{\bar n(z-1)}{k} \right] ^{-k} 
\ee
\be
G_{NB}(\bar n,k;z) = 
G_{Poisson} [\bar N; G_{log}(\bar n,k;z)]= \exp \left[\bar N 
\left( G_{log} - 1 \right)\right] 
\ee                   
with  $\bar N= k \log ( 1 + \bar n/k)$ and  $\bar n_c = \bar n/ \bar N$,
the average number of clan and the average number of particles per clan 
respectively, i.e, $G_{NB}(\bar n,k;z)$ 
is a Compound Poisson Distribution (CPD) or discrete infinitely
divisible   distribution.
At this level
the introduction of the clan concept  is  essentially a change in the
variables one suggests to observe in the multiplicity distribution, i.e.
\be
\hbox{from} \qquad {\bar n \choose k} \qquad \hbox{to} \qquad  
{\bar N \choose \bar n_c} 
\label{clan}
\ee

Notice that  $1/k$  is interpreted  here as a measure of aggregation of
particles into clans
\be
\frac{1}{k} = \frac{P_1(2)}{P_2(2)}
\ee
i.e., it corresponds to the ratio of the probability  to have two particles
in the same clan, $P_1(2)$, over  the probability  to have the two particles
into two separate clans, $P_2(2)$.

Finally it should be recalled that in more general terms  $1/k$ is linked
to the two particle rapidity correlations $C_2(y_1,y_2)$  and to the 
second order factorial cumulant,  $\kappa_2$, by the equation
\be
\frac{1}{k}= \kappa_2 =\int C_2(y_1,y_2) dy_1 dy_2
\ee
The  deep connection among clan structure parameters and correlations
is starting from these very simple remarks.

The analysis of MD's in terms of the new parameters (clan structure analysis)
reveals new striking regularities: 

a.  linear increase of $\bar N$ with the rapidity cut, $y_c$, at fixed  c.m. 
energy; as $y_c$ approaches full phase space,  $\bar N$ is bending towards a
constant value; 

b.  energy independence of $\bar N$ 
in a fixed rapidity interval in the region of linearity; 

c.  $\bar N$ is larger in $e^+e^-$ than in $hh$ and in $lh$ reactions
  whereas $\bar n_c$ is smaller in $e^+e^-$ annihilation and $lh$ than
  in $hh$   reactions.

\subsection{NB regularity in Monte Carlo calculations 
(La Thuile Symposium  1989 and then Perugia and Ringberg  Workshops)}
In order to have  informations  on MD's at final parton level at 
different c.m. energies   
in full phase space and in rapidity intervals , 
i.e., in a region where QCD predictions are poor, 
clan  structure analysis has been applied to $q\bar q$ and $gg$ systems
as resulting in JETSET Monte Carlo calculations. Results of this search
have been presented at the la Thuile Symposium in 1989  and in the same
period at the Perugia and Ringberg Workshops. 
This study shows:

a. final partons  and charged hadron multiplicities in full phase
space and in symmetric rapidity intervals are well fitted by
NB MD's.

b. Hadronic and partonic
levels are not independent  but linked by the following equations
\be
\frac{1}{k} \biggl|_{hadron} = \frac{1}{k} \biggl|_{parton} 
\quad , \quad 
\bar n \biggl|_{hadron} = \rho \bar n \biggl|_{parton} 
\ee
with $\rho$ a constant $\sim 2Q_0/$1 GeV, $Q_0$ is the parton 
virtuality cut-off, 
or, in terms of $n$-particle (parton) inclusive multiplicity distributions
\be
Q_n \biggl|_{hadron} (y_1,\dots,y_n) =  \rho^n 
Q_n \biggl|_{parton} (y_1,\dots,y_n)
\ee
which can be considered a generalization of preconfinement and defines
the Generalized Local Parton Hadron Duality (GLHPD) hadronization 
prescription.

c. The general trend of the average number of clans is at both levels
consistent with what was previously discovered by clan structure analysis
at hadron level.In addition the regularity is more impressive at parton 
level than at hadron level, i.e., where Altarelli-Parisi  or 
Konishi-Ukawa-Veneziano QCD evolution  equations predictions
are not affected by the Monte Carlo hadronization  prescription.

\subsection{The shoulder structure puzzle and NB regularity in jets of fixed
topology (La Thuile 1989, Moriond 1991, 
Wuhan 1991, Vietri 1994, Stara Lesna 1995)} 

As already mentioned in $p\bar p$ reactions UA5
collaboration noticed that MD of the full sample of events 
at c.m. energy $\sqrt{s}$ = 900 GeV  in largest
pseudorapidity intervals violates  NB regularity and shows a characteristic
shoulder  structure. The effect was reported by Fuglesang at the La Thuile
Symposium. Later the same effect was also seen at LEP  at $\sqrt{s}$ =
91 GeV.

The conclusion of those  who were afraid of the regularity was 
that NB regularity was dead.

At the same time more conservative people asked to look better at the 
data: shoulder effect might very well be due  to  the overlap of jets of
different topologies  i.e. the regularity 
should be tested at a more elementary level.
The results of this new analysis  in $e^+e^-$ at LEP energy
 are presented at the XXVI Rencontres
de Moriond in 1991 by V. Uvarov. Although the analysis is dependent
on the jet selection algorithm (JADE algorithm was used) three 
important facts occur: 

\noindent 
a. the shoulder in $e^+e^-$ annihilation is the result of the 
superposition of jets of different topologies;

\noindent 
b. MD's in jets of different topologies are nicely fitted by NB MD's, 
i.e., the regularity which is violated in the full sample
of events is approximately restored  at a more elementary level 
of investigation; 

\noindent 
c. LEP data less approximate on sample of two jet events 
are fully consistent with HRS  Collaboration data at lower c.m. energy.

In particular clan structure analysis reveals that standard trends for 
$\bar N$ and $\bar n_c$ are less approximate   on the sample of two jet
events than in the full sample.
In addition by applying GLHPD backwards 
one sees new impressive regularities at parton level: the average 
number of clan at parton level  is 1 in $\Delta y$ =1, 2 in 
$\Delta y$=2, $\dots$, 5 in $\Delta
y$=5. This fact is verified both at LEP and HRS  c.m. energies. 
The average number of clans is a linear function of rapidity variable.
The average number of partons (particles) per clan follows the standard
trend.
It should be stressed that the same mechanism, i.e., the superposition of 
jets of different topologies (hard gluon radiation) explains in part 
also $H_q$ oscillations.

\subsection{New  results (Faro 1996)}

Two new results have been presented at this Symposium.
a. $H_q$ oscillations and NB regularity in quark jets of fixed flavour.
 In  $e^+e^-$ annihilation at the  $Z^0$  peak   
either the charged particle MD,  $P_n$, or $H_q$ oscillations are well 
reproduced in  full phase space  by the weighted superposition of two 
NB MD's associated to $b\bar b$ and light flavoured events, the 
weight being
given by the fraction of $b\bar b$ events. The corresponding NB parameters
are characterized by the fact that\cite{nostri}
\be
\bar n_{\hbox{\scriptsize\it light flavoured events}} \ne
 \bar n_{b\bar b~\hbox{\scriptsize\it events}}
\quad , \quad 
k_l = k_b
\ee
Clan structure analysis suggests   that the aggregation of particles into
clan is the same  in  $b\bar b$ and light flavoured events!

\noindent 
b. clan structure analysis at H1 experiment at HERA.
H1 Collaboration at HERA presented data on MD's  in different 
pseudorapidity windows 
 at various energy intervals  and then  fitted these data
 by NB MD\cite{h1}.
Clan structure analysis performed on these data  reveals  that the average 
number of clans 
in a fixed rapidity interval is energy independent and that the 
average number
of particle per clan  follows the same general trend already observed
in EMC data (the figure presented at this Symposium is not shown here
due to the lack of space).

\section{Theory}

\subsection{Jets as QCD Markov branching processes  (Goa 1979, Bruges 1980, 
Wuhan 1991, Vietri 1994)} 

QCD jets have been described  in LLA as Markov branching processes of
NB type since 1979\cite{AGQCD}. 
The  approach is a purely exclusive one and consists in solving
explicitly KUV equations  with a fixed cut off regularization 
prescription, once the 
intrinsic Markoffian structure of the process has been recognized.
It should be recalled that    
KUV equations   are deduced in an inclusive
approach to the production process and are the differential formulation of
 the  integral AP QCD  equations\cite{KUV}. 
Results of this search have been extensively discussed in Goa (1979),
Bruges (1980), Wuhan (1991) and Vietri (1994) Symposia. 
The main aspect of this
study is that parton showers behave here as self-similar fractals; 
they develop according to a principle of {\it disorder} (fractal
from Latin {\it frangere}, to break) followed  by a 
principle of {\it order} 
and controlled by self-similarity, i.e., according to 
an "Italian firework" of jets of jets... In other words  
in a quark parton generated shower the ancestor quark "breaks" into 
 bremsstrahlung gluon jets (the  clans)
which then decay into final parton via gluon self interaction, 
and partons belonging to each clan are distributed
according to a weighted geometric
distribution. What is interesting to point out is that 
clan structure analysis decouples the   $g \to g+g$ vertex 
from the $q \to q+g$ vertex  and that $1/k$ is here
just their ratio.

\subsection{Clan structure analysis  in statistical physics 
(Aspen 1993, Vietri 1994)}

This study concerns: 

\noindent 
a. clans and compound Poisson distributions\cite{void}; 
b. clans and hierarchical correlation functions\cite{lpa}.  
It has been done in collaboration with Sergio Lupia and Roberto Ugoccioni
and reported in Aspen 1993 and Vietri 1994 Symposia.

a. The description of multiparticle production in a  domain 
of rapidity $\Delta y$  as a two step process
can be formulated in very general terms in the framework of infinitely
divisible distributions (IDD).  Any discrete IDD  
is indeed a Compound Poisson Distribution  (CPD).
In the generating functions formalism this important property leads
in a natural way to the generalized clan concept (g-clan) as can be seen by 
inspection of
the following general definition of the generating function of any CPD: 
\be
G_{CPD}(z;\Delta y) = \exp \left[ \bar N_g (\Delta y) [ G'(z;\Delta y) -
1]\right]     
\ee
where $\bar N_g$ is the average number of g-clans and $G'(z;\Delta y)$ 
is the generating
function of the MD inside an average g-clan. Since by definition each g-clan 
contains  at least one particle the condition $G'(z;\Delta y) |_{z=0}$ = 0
has to be imposed. 
Accordingly clan concept is more general than NB multiplicity 
distribution, whose
wide occurrence in final particle multiplicity distributions 
justified its first
 introduction. In addition a simple relation exists for any CPD
among the probability  $P_0(\Delta y)$ to
detect no particle in a given rapidity interval $\Delta y$ and the average
number of generalized clan, $\bar N_g$, i.e.
\be
G_{CPD}(0;\Delta y) = e^{- \bar N_g(\Delta y)} = P_0(\Delta y)     
\ee

b. The main result   can be  summarized  by the following
theorem: factorial cumulant structure is hierarchical , i.e., 
$n$-order factorial  cumulants, $\kappa_n(\Delta y)$,  
are controlled by  second order factorial cumulants, 
$\kappa_2 (\Delta y)$, only 
{\it iff} the  function  
${\cal V}(\Delta y) \equiv -\log P_0(\Delta y)/\bar n(\Delta y)$  
scales with energy and $\Delta y$ 
as a function of $\bar n(\Delta y) \kappa_2(\Delta y)$. 
Notice that the just mentioned 
scaling function is the inverse 
of the average number of particles per generalized clan, $(\bar n_c)_g$. 

In conclusion clan structure parameters can be extended 
to the  class of discrete IDD 's
(they are the most economic way to 
describe a two step production process in which
independently generated extended 
objects -- clans, clusters, strings, $\ldots$ -- 
decay into
final particles or partons). In addition these parameters 
are useful and intriguing in view of their
deep physical meaning (they control "voids" distribution 
in rapidity and being 
$\kappa_n(\Delta y) = \int_{\Delta y} dy_1 \dots \int_{\Delta y} dy_n$   
$c_n(y_1,\dots,y_n)$  the hierarchical
content of $n$-particle (-parton) 
correlation functions  $c_n(y_1,\dots,y_n)$  
in the production process).
Therefore the scheme described in~\eref{clan} 
should be completed  as follows
\be
\hbox{from} \qquad {\bar N(\Delta y) \choose \bar n_c(\Delta y)} 
\qquad \hbox{to} \qquad  {P_0(\Delta y) \choose {\cal V}(\Delta y)} 
\ee
In case of the NB MD the new set of parameters is linked  
to the previous one by the equations
\be   
 \bar N(\Delta y) = - \log P_0(\Delta y) = - \log  
\left[ \frac{k(\Delta y)}{\bar n(\Delta y) + 
k(\Delta y)} \right]^{k(\Delta y)}
\ee
\be
\bar n_c(\Delta y) = 
\frac{\bar n(\Delta y) \kappa_2(\Delta y)}{\log [1+ \bar n(\Delta y)
\kappa_2(\Delta y)]} =  {\cal V}(\Delta y)^{-1}
\ee                                                      
The second equation shows that  NB regularity predicts that 
$n$-particle correlations are hierarchical\cite{lpa}. 

\subsection{Clan structure parameters in theoretical models 
(Vietri 1994, Stara Lesna 1995)}

The energy dependence of
clan structure parameters  can be calculated analytically  either in f.p.s.
or in rapidity intervals in the generalized simplified parton shower model
(GSPS)\cite{GSPS}.
The model is a generalization of the work initiated with Leon Van
Hove\cite{SPS}. It
is  a two parameter model and is based on {\it essentials of QCD}  
(gluon self interaction) in a convenient kinematical framework: 
energy- momentum conservation laws are violated
locally but not globally. Clans are  considered 
here as independent intermediate fluctuating
gluon sources. Observed general trends of clan 
structure parameter in energy and rapidity
are correctly predicted\cite{GSPS,GSPS2}. These results suggest that
the ancestors of the clans, i.e.,  
the independently emitted bremsstrahlung gluons, 
 should be indeed identified at parton
level with the just mentioned fluctuating gluon sources.

\subsection{Clans and the hadronization problem (Stara Lesna 1995, Faro
1996)}

At this point we are ready to ask the following question:
 are clans observable physical objects?
A possible hint comes from the thermodynamical hadronization
model\cite{becattini}. The model has been successfully applied 
to hadron production 
in $e^+e^-$ annihilation. The hadronic spectrum is quite well 
reproduced in terms
 of three parameters (the temperature, the volume and 
strangeness chemical suppression).
The production process is described  as a two-step process  in which 
primary hadrons emitted by the thermal source decay into final particles. 
It is interesting to
point out that final charged tracks MD's turn out to be of NB type and the
average number of clans calculated from fitted NB coincides with the 
average number of primary hadrons predicted by the thermodynamical 
model\cite{bgl}. 
This result suggests that at hadron level in $e^+e^-$ annihilation 
clans should be identified with primary hadrons.

\section{Conclusions}

The  main results of experimental and theoretical work 
done along the years on approximate
NB regularity and related clan structure, and reported in 
Multiparticle Dynamics Symposia, have  been summarized. 
The hope is to have given a perspective for future work
and a contribution to the answer of the following questions.

Is NB  regularity still an enigma -- 
as Leon Van Hove pointed out in Shandong in 1987 -- 
or a useful Arianna's thread -- as I prefer now -- in order to drive us  
in the labyrinth of multiparticle dynamics?

Is the fact that  in $e^+e^-$ annihilation NB MD can be associated to jets
of different flavour accidental, or does it reveals the 
intimate structure of
multiparticle dynamics in the building block of the reaction? 
If it is so,  what about other reactions?

Is clan concept a purely statistical one (like cluster expansion in 
statistical mechanics) or are clans observable physical objects?

Are clans at parton level bremsstrahlung gluon jets? 
Can one consider the ancestors of
each clan (the bremsstrahlung gluons) independently emitted
 fluctuating gluon sources?

Are really clans at hadron level in $e^+e^-$  annihilation primary hadrons?
If it is so, what about other reactions?

\section*{Acknowledgements} 
I thank Jorge Dias de Deus %and the whole Organizing Committee  
for the nice atmosphere created at the Conference. 
I  also wish to express my gratitude to Sergio Lupia and Roberto
Ugoccioni for a critical reading of the manuscript.

\end{document}